\documentclass[reprint,amsmath,amssymb,aps,prd,noeprint,nolongbibliography]{revtex4-2}
\usepackage[utf8]{inputenc}
\usepackage{graphicx}
\usepackage{dcolumn}
\usepackage{bm}
\usepackage[colorlinks,citecolor=blue,urlcolor=blue,linkcolor=blue]{hyperref}
\usepackage[mathlines]{lineno}
\usepackage{color}
\usepackage{slashed}
\usepackage{multirow}
\usepackage{makecell}
\usepackage{booktabs}
\usepackage{tabularx}
\usepackage[symbol]{footmisc}
\usepackage{comment}

\newcommand{\gev}{\ensuremath{\mathrm{\,Ge\kern -0.1em V}}\xspace}
\newcommand{\mev}{\ensuremath{\mathrm{\,Me\kern -0.1em V}}\xspace}
\newcommand{\mevcc}{\ensuremath{{\mathrm{\,Me\kern -0.1em V\!/}c^2}}\xspace}

\usepackage{lineno}
\usepackage{hyperref}
\hypersetup{hypertex=true,
	        colorlinks = true,
            linkcolor = blue,
            urlcolor = blue,
            citecolor = blue}

\begin{document}

\newcommand{\BESIIIorcid}[1]{\href{https://orcid.org/#1}{\hspace*{0.1em}\raisebox{-0.45ex}{\includegraphics[width=1em]{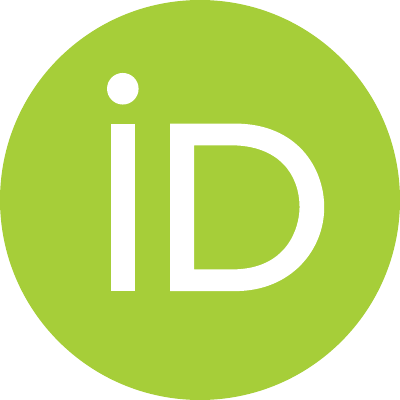}}}}

\title{\boldmath Observation of the $\chi_{cJ}$ decays into $pK^{-}\bar{\Lambda}\eta+\mathrm{c.c.}$ }

\author{
M.~Ablikim$^{1}$\BESIIIorcid{0000-0002-3935-619X},
M.~N.~Achasov$^{4,c}$\BESIIIorcid{0000-0002-9400-8622},
P.~Adlarson$^{84}$\BESIIIorcid{0000-0001-6280-3851},
X.~C.~Ai$^{90}$\BESIIIorcid{0000-0003-3856-2415},
C.~S.~Akondi$^{32A,32B}$\BESIIIorcid{0000-0001-6303-5217},
R.~Aliberti$^{40}$\BESIIIorcid{0000-0003-3500-4012},
A.~Amoroso$^{83A,83C}$\BESIIIorcid{0000-0002-3095-8610},
Q.~An$^{79,66,\dagger}$,
Y.~H.~An$^{90}$\BESIIIorcid{0009-0008-3419-0849},
M.~S.~Anderson$^{40}$\BESIIIorcid{0009-0008-1550-2632}
% ...
\\
\vspace{0.2cm}
(BESIII Collaboration)\\
\vspace{0.2cm} {\it
$^{1}$ Institute of High Energy Physics, Beijing 100049, People's Republic of China\\
% ...
$^{4}$ Budker Institute of Nuclear Physics SB RAS (BINP), Novosibirsk 630090, Russia\\
% ...
$^{32}$ INFN Sezione di Ferrara, (A)INFN Sezione di Ferrara, I-44122, Ferrara, Italy; (B)University of Ferrara, I-44122, Ferrara, Italy\\
% ...
$^{40}$ Johannes Gutenberg University of Mainz, Johann-Joachim-Becher-Weg 45, D-55099 Mainz, Germany\\
% ...
$^{66}$ State Key Laboratory of Particle Detection and Electronics, Beijing 100049, Hefei 230026, People's Republic of China\\
% ...
$^{79}$ University of Science and Technology of China, Hefei 230026, People's Republic of China\\
% ...
$^{83}$ University of Turin and INFN, (A)University of Turin, I-10125, Turin, Italy; (B)University of Eastern Piedmont, I-15121, Alessandria, Italy; (C)INFN, I-10125, Turin, Italy\\
$^{84}$ Uppsala University, Box 516, SE-75120 Uppsala, Sweden\\
% ...
$^{90}$ Zhengzhou University, Zhengzhou 450001, People's Republic of China\\
\vspace{0.2cm}
$^{\dagger}$ Deceased\\
% ...
$^{c}$ Also at the Novosibirsk State University, Novosibirsk, 630090, Russia\\
% ...
}
}

\begin{abstract}
By analyzing $(2712.4 \pm 14.3) \times 10^{6}$ $\psi(3686)$ events collected with the
BESIII detector operating at the BEPCII collider, the decays
$\chi_{cJ} \to pK^{-}\bar{\Lambda}\eta + \mathrm{c.c.}$ ($J=0,1,2$) are observed
for the first time, with statistical significances exceeding $5\sigma$ for all three $\chi_{cJ}$ states. The measured branching fractions are $\mathcal{B}(\chi_{c0} \to pK^{-}\bar{\Lambda}\eta + \mathrm{c.c.})
= (5.3 \pm 0.7 \pm 0.5) \times 10^{-5}$,
$\mathcal{B}(\chi_{c1} \to pK^{-}\bar{\Lambda}\eta + \mathrm{c.c.})
= (9.8 \pm 0.6 \pm 0.6) \times 10^{-5}$,
and $\mathcal{B}(\chi_{c2} \to pK^{-}\bar{\Lambda}\eta + \mathrm{c.c.})
= (9.3 \pm 0.6 \pm 0.6) \times 10^{-5}$, where the first uncertainties are statistical and the second are systematic.
Structures consistent with the known hyperon resonances $\Lambda(1520)$ and
$\bar\Lambda(1690)$ are seen in the $pK^{-}$ and $\bar{\Lambda}\eta$ invariant mass spectra, respectively. The reported branching fractions include both resonant and non-resonant contributions.
These results provide new experimental information on hadronic decays of $P$-wave charmonium states and contribute to the understanding of baryon production and hadronization dynamics in the nonperturbative QCD regime.
\end{abstract}

\maketitle

\section{Introduction}
In the quark model, the $\chi_{cJ}$ states ($J = 0, 1, 2$) correspond to the $^3P_J$ charmonium states.
Their quantum numbers are incompatible with those of a single virtual photon, and therefore they cannot be produced through single-photon exchange in $e^+e^-$ annihilation~\cite{BESIII:2022mtl}.
The $\chi_{cJ}$ states are efficiently produced via the radiative transitions $\psi(3686)\to\gamma\chi_{cJ}$, whose branching fractions are about 9\% for each $\chi_{cJ}$ state~\cite{ParticleDataGroup:2024cfk}.
These transitions make $\psi(3686)$ decays an excellent laboratory for studying the hadronic decay properties of $\chi_{cJ}$ states and for testing the predictions of nonperturbative quantum chromodynamics (QCD)~\cite{Bodwin:1994jh}. Hadronic decays of $\chi_{cJ}$ proceed predominantly through the annihilation of a $c\bar{c}$ pair into gluons, followed by hadronization into multi-hadron final states. Such processes are sensitive to intermediate resonant structures and final-state interactions, thereby providing valuable information on hadronization dynamics and baryon production mechanisms~\cite{QuarkoniumWorkingGroup:2004kpm}.

In the previous studies at the BES and BESIII experiments,
a near-threshold enhancement in the $p\bar{\Lambda}$ invariant-mass spectrum
was observed in decay channels of $J/\psi$ and $\psi(3686)$~\cite{BES:2004fgd}, and similar behavior was also reported in $\chi_{cJ}\to pK^{-}\bar{\Lambda}$ decays~\cite{BESIII:2012koo}. This structure has attracted considerable interest in hadron spectroscopy, and has been interpreted in terms of possible baryonium-like states~\cite{Deng:2013aca}, strong final-state interactions~\cite{Haidenbauer:2024smo}, or interference effects from excited kaon contributions~\cite{Wang:2020wap}. Extending such investigations to more $\chi_{cJ}$ decay modes can help further
clarify the role of similar effects in $P$-wave charmonium transitions, where
the gluon hadronization dynamics may differ from those in $S$-wave decays.

Given the considerations discussed above, the decays $\chi_{cJ} \to pK^{-}\bar{\Lambda}\eta$
provide a unique opportunity to study baryon--antibaryon threshold behavior and hyperon excitation dynamics. The final state contains both a baryon--antibaryon pair and pseudoscalar mesons, making it particularly sensitive to potential excited $\Lambda$ hyperon contributions in the $pK^{-}$ and $\bar{\Lambda}\eta$ subsystems. In this paper, by analyzing $(2712.4 \pm 14.3)\times10^{6}$ $\psi(3686)$ events collected with the BESIII detector~\cite{BESIII:2024lks}, we present the first observation and of the decays $\chi_{cJ} \to pK^{-}\bar{\Lambda}\eta$ and measure their corresponding branching fractions. Charge-conjugate modes are implied throughout this paper.

\section{BESIII DETECTOR AND MONTE CARLO SIMULATION}
\label{sec:BES}
The BESIII detector~\cite{BESIII:2009fln} records symmetric $e^+e^-$ collisions provided by the BEPCII storage ring~\cite{Yu:2016cof}
in the center-of-mass energy range from 1.84 to 4.95~GeV,
with a peak luminosity of $1.1 \times 10^{33}\;\text{cm}^{-2}\text{s}^{-1}$
achieved at $\sqrt{s} = 3.773\;\text{GeV}$.
The cylindrical core of the BESIII detector covers 93\% of the full solid angle and consists of a helium-based
 multilayer drift chamber~(MDC), a plastic scintillator time-of-flight
system~(TOF), and a CsI(Tl) electromagnetic calorimeter~(EMC),
which are all enclosed in a superconducting solenoidal magnet
providing a 1.0~T magnetic field, except for the data taken in 2012,
for which the magnetic field was 0.9~T. The solenoid is supported by an
octagonal flux-return yoke with resistive plate counter muon
identification modules interleaved with steel.
The charged-particle momentum resolution at $1~{\rm GeV}/c$ is
$0.5\%$, and the
${\rm d}E/{\rm d}x$
resolution is $6\%$ for electrons
from Bhabha scattering. The EMC measures photon energies with a
resolution of $2.5\%$ ($5\%$) at $1$~GeV in the barrel (endcap)
region. The time resolution in the TOF barrel region is 68~ps, while
that in the endcap region was 110~ps.
The endcap TOF system was upgraded in 2015 using multi-gap resistive plate chamber technology, providing a time resolution of 60~ps~\cite{Li:2017jpg,Guo:2017sjt,Cao:2020ibk}. Around 83\% of the data used in this analysis benefits from this upgrade.

Simulated data samples produced with a {\sc
geant4}-based~\cite{GEANT4:2002zbu} Monte Carlo (MC) package, which
includes the geometric description~\cite{Huang:2022wuo,Li:2024pox,Liang:2009zzb,Ya-Jun:2008mrr} of the BESIII detector and the
detector response, are used to determine detection efficiencies
and to estimate backgrounds. The simulation models the beam
energy spread and initial-state radiation (ISR) in the $e^+e^-$
annihilations with the generator {\sc
kkmc}~\cite{Jadach:2000ir}. An inclusive MC sample containing $2.7\times10^{9}$ generic $\psi(3686)$ events is used to study potential backgrounds. The inclusive MC sample includes the production of the
$\psi(3686)$ resonance.
Known particle decays are modeled with {\sc
evtgen}~\cite{Lange:2001uf,Ping:2008zz} using branching fractions
taken from the
Particle Data Group (PDG)~\cite{ParticleDataGroup:2024cfk}, while unknown decays are modeled with {\sc lundcharm}~\cite{Chen:2000tv}.
Final-state radiation
from charged final state particles is incorporated using the {\sc
photos} package~\cite{Richter-Was:1992hxq}.

To account for the effect of intermediate resonance structures on the detection efficiency, each of the $\chi_{c0,1,2}$ decays is modeled using 5.04 million mixed signal MC samples.
The dominant decay modes include intermediate resonances such as $\Lambda(1520)$ and $\bar\Lambda(1690)$, corresponding to the $pK^-$ and
$\bar{\Lambda}\eta$ systems, respectively. These components are generated using a phase-space (PHSP) model employed in {\sc
evtgen}.
These samples are then mixed with four-body PHSP signal MC samples.
The mixing ratios are determined by examining the corresponding invariant mass spectra, as discussed in Sec.~\ref{sec:bf}.

\section{EVENT SELECTION}
\label{sec:selection}
We study the process $\psi(3686)\to\gamma\chi_{cJ}$ followed by
$\chi_{cJ}\to pK^{-}\bar\Lambda\eta$, with
$\bar\Lambda\to \bar p\pi^{+}$ and
$\eta\to\gamma\gamma$.
Charged tracks ($p$, $K^{-}$, $\bar p$, and $\pi^{+}$) and photons are reconstructed with the standard BESIII algorithms.
Candidate events are required to have at least four charged tracks with zero net charge.
Tracks reconstructed in the MDC must satisfy
$\vert\!\cos\theta\vert<0.93$, where $\theta$ is the polar angle with respect to
the MDC $z$ axis.
For charged tracks originating directly from the $\chi_{cJ}$ decay, the distances of closest approach to the interaction point (IP) are required to satisfy
$|V_z|<10$~cm and $|V_{xy}|<1$~cm,
where $V_z$ and $V_{xy}$ denote the distances along the beam and transverse directions, respectively.
No IP-vertex requirement is imposed on the daughter tracks from the $\bar\Lambda$ decay.

The PID information combines the 
${\rm d}E/{\rm d}x$
measurements in the MDC and the TOF information to form likelihoods $\mathcal{L}(h)$ for
$h\in\{p,\,K,\,\pi\}$.
A track is identified as a proton if
$\mathcal{L}(p)>\mathcal{L}(K)$ and $\mathcal{L}(p)>\mathcal{L}(\pi)$,
and as a kaon if
$\mathcal{L}(K)>\mathcal{L}(p)$ and $\mathcal{L}(K)>\mathcal{L}(\pi)$.
Tracks failing both criteria are treated as pion candidates.

Photon candidates are identified using isolated showers in the EMC.
At least three good photon candidates are required.
The deposited energy must exceed 25~MeV in the barrel region ($\vert\!\cos\theta\vert<0.80$) and 50~MeV in the endcap region ($0.86<\vert\!\cos\theta\vert<0.92$).
The EMC time is required to be within $[0,700]$~ns relative to the event start time to suppress noise and beam backgrounds.
To reduce contamination from hadronic interactions or bremsstrahlung, clusters within a cone of $20^\circ$ around the extrapolated EMC position of a $\bar p$ track and within $10^\circ$ of any other charged track are rejected.

\begin{figure}[htbp]     \centering     \includegraphics[width=0.45\textwidth]{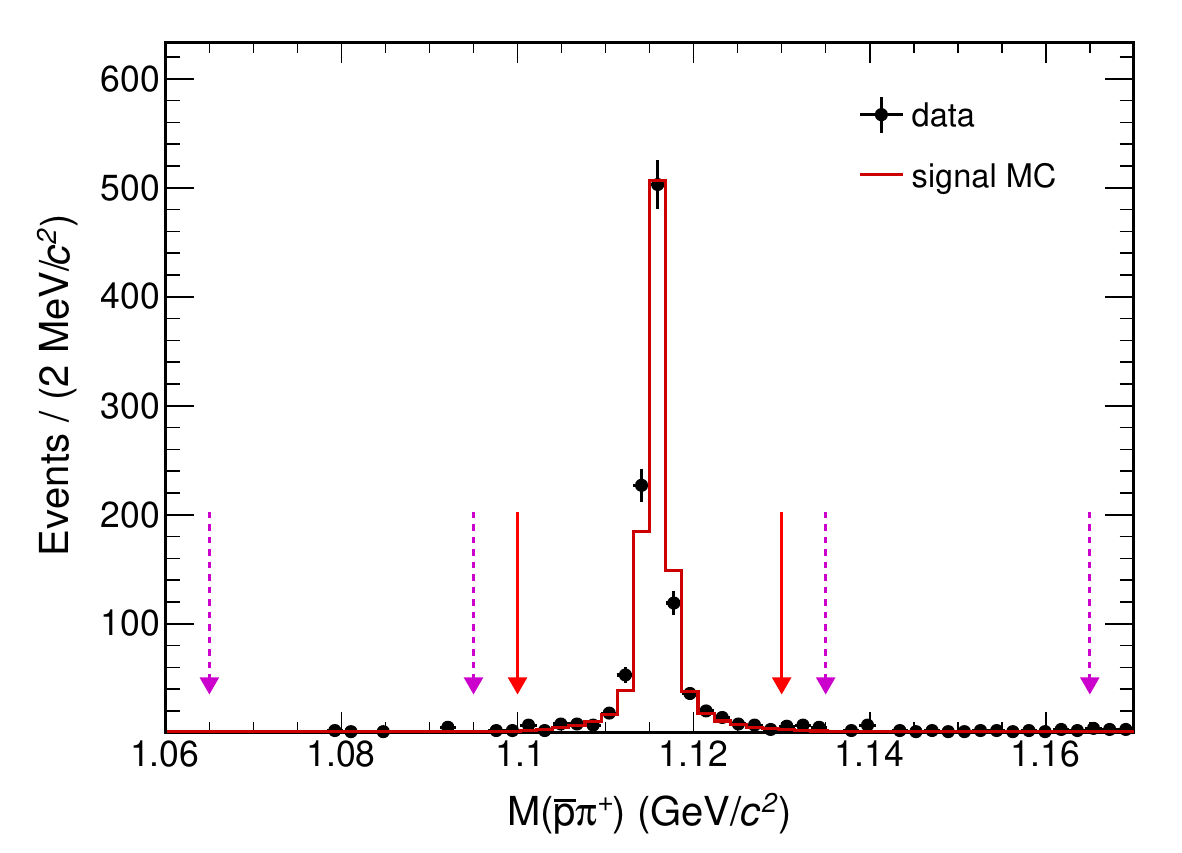}     
\caption{Invariant mass spectrum of  $\bar p\pi^+$ candidates.  Data are shown as black points with
error bars, and the signal MC distribution is shown as the red histogram. The
$\bar\Lambda$ signal window [1.100, 1.130]~GeV$/c^2$ is indicated by
red solid arrows, and the sideband windows [1.065, 1.095]~GeV$/c^2$ (left) and
[1.135, 1.165]~GeV$/c^2$ (right) are indicated by magenta dashed arrows.}
 \label{fig:lambda_mass}
 \end{figure}

The $\bar\Lambda$ candidates are reconstructed from pairs of oppositely charged tracks excluding the selected proton and kaon candidates, under the
$\bar p\pi^{+}$ mass hypothesis, without applying PID requirements to the daughter
tracks. The two tracks are required to satisfy $\vert\!\cos\theta\vert<0.93$ and $|V_{z}|<20$~cm; no
requirement is imposed on $V_{xy}$ due to the displaced decay vertex of the long-lived $\bar\Lambda$ baryon.
The two tracks are subjected to a common secondary-vertex fit, and combinations with
fit $\chi^{2}_{\rm vtx}<200$ are retained. To suppress combinatorial background, the decay length of the $\bar\Lambda$ candidate is required to satisfy
$L/\sigma_L>2$, where $L$ is the distance between the reconstructed decay vertex and the IP, and $\sigma_L$ is the uncertainty of the decay length $L$.
The $\bar\Lambda$ signal region is defined as
$|M_{\bar p\pi^{+}}-m_{\bar\Lambda}|<0.015$~GeV$/c^{2}$,
while the sidebands are
$0.020<|M_{\bar p\pi^{+}}-m_{\bar\Lambda}|<0.050$~GeV$/c^{2}$, where $m_{\bar\Lambda}$ is the known $\bar\Lambda$ mass~\cite{ParticleDataGroup:2024cfk};
the distributions and windows are shown in Fig.~\ref{fig:lambda_mass}.

To improve the mass resolution and suppress combinatorial background,
a four-constraint (4C) kinematic fit is performed under the hypothesis
$e^{+}e^{-}\!\to\!pK^{-}\bar{\Lambda}\gamma\gamma\gamma$, imposing conservation of the
total energy and the three-momentum components.
If multiple combinations are found, the one with the smallest
fit $\chi^{2}_{\rm 4C}$ is retained.  
The requirement on $\chi^{2}_{\rm 4C}$ is optimized using a figure-of-merit (FOM) procedure, 
where the FOM is defined as
${\rm FOM}=S/\sqrt{S+B}$ with $S$ and $B$ denoting the signal and background yields extracted from the signal MC and inclusive MC samples normalized to data. 
This yields the nominal selection $\chi^{2}_{\rm 4C}<35$.

\begin{figure}[htbp]
    \centering
 \includegraphics[width=0.45\textwidth]{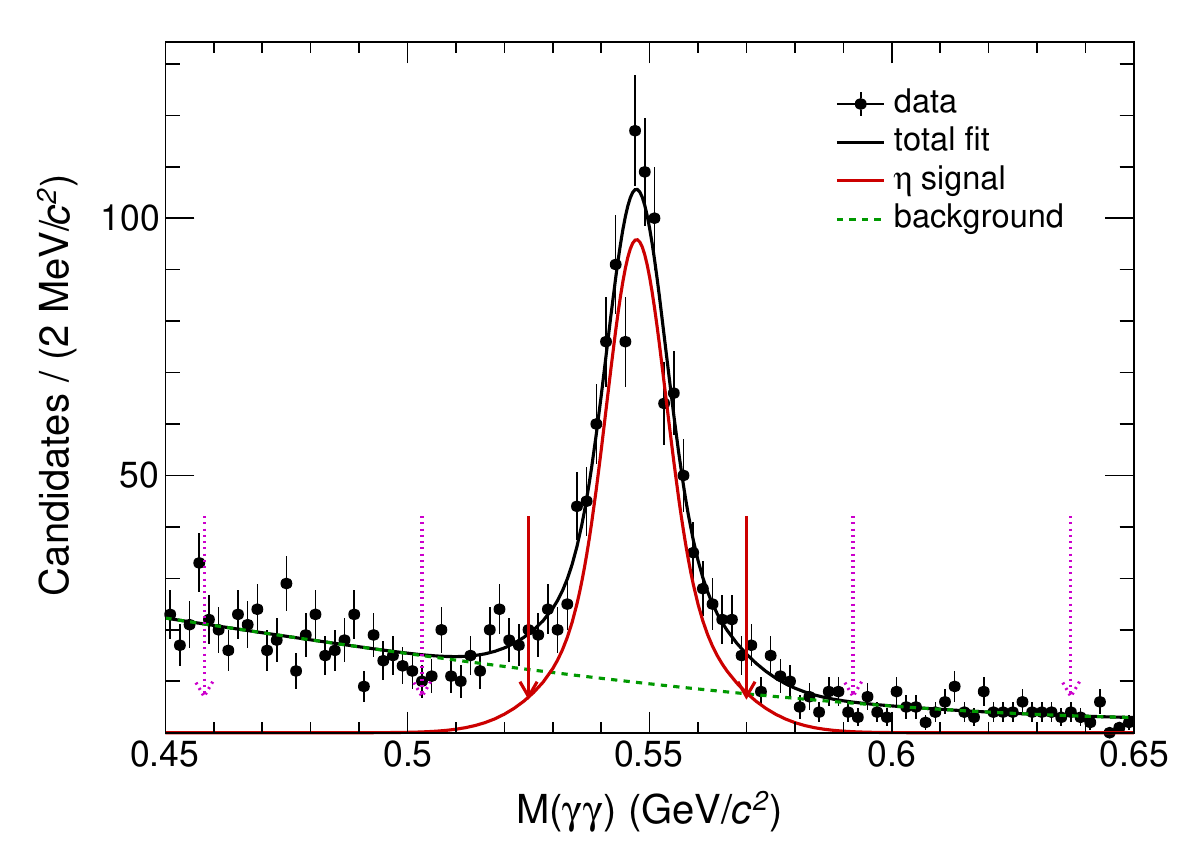}
    \caption{Invariant mass spectrum of $\gamma\gamma$ combinations for $\eta$ candidates in data. The total fit is shown by the black solid curve, while the red solid curve represents the $\eta$ signal component and the green dashed curve denotes the background contribution. The $\eta$ signal region $[0.525,\,0.570]~\mathrm{GeV}/c^{2}$ is indicated by red solid arrows, 
and the sideband regions $[0.458,\,0.503]~\mathrm{GeV}/c^{2}$ (left) and 
$[0.592,\,0.637]~\mathrm{GeV}/c^{2}$ (right) are indicated by magenta dashed arrows.
}

    \label{fig:eta_mass}
\end{figure}

The $\eta$ meson candidate is reconstructed from photon pairs formed from the three selected photons. The three photons are ordered by energy, $E_{\gamma_1}>E_{\gamma_2}>E_{\gamma_3}$. Studies with signal MC indicate that the $(\gamma_2\gamma_3)$ combination is found to contain a negligible $\eta$ signal and is therefore discarded; the $\eta$ candidate is formed from either the $\gamma_1\gamma_2$ or $\gamma_1\gamma_3$ combination. The $M_{\gamma\gamma}$ distribution is fitted with a double-Gaussian signal function and a polynomial background function, as shown in Fig.~\ref{fig:eta_mass}. The $\eta$ signal window is defined as $0.525<M_{\gamma\gamma}<0.570~\mathrm{GeV}/c^{2}$, and the lower (upper) sideband region is defined as $0.458<M_{\gamma\gamma}<0.503~\mathrm{GeV}/c^{2}$ ($0.592<M_{\gamma\gamma}<0.637~\mathrm{GeV}/c^{2}$). If both $M_{\gamma_1\gamma_2}$ and $M_{\gamma_1\gamma_3}$ fall within the $\eta$ signal window in the same event, the event is excluded from further analysis to avoid double counting, at the loss of about 4\% of the $\eta$ signal yield.

To suppress $\pi^{0}\!\to\!\gamma\gamma$ contamination, photon-pair veto
windows around the known $\pi^{0}$ mass $m_{\pi^{0}}$~\cite{ParticleDataGroup:2024cfk} are applied, 
$|M_{\gamma_{1}\gamma_{2}}-m_{\pi^{0}}|>0.013~\mathrm{GeV}/c^{2}$,
$|M_{\gamma_{1}\gamma_{3}}-m_{\pi^{0}}|>0.018~\mathrm{GeV}/c^{2}$,
and $|M_{\gamma_{2}\gamma_{3}}-m_{\pi^{0}}|>0.009~\mathrm{GeV}/c^{2}$, due to the different $\pi^{0}$ invariant-mass resolutions of the three photon combinations.
These requirements, together with all other veto selections described
below, are optimized using the same FOM procedure, based on the signal and inclusive MC samples nor-malized to data.
Backgrounds from $\psi(3686)\!\to\!\eta J/\psi$ with
$J/\psi\!\to\!pK^{-}\bar{\Lambda}+\mathrm{c.c.}$ are suppressed by vetoing events whose
$\eta$ recoil mass lies near the known $J/\psi$ mass $m_{J/\psi}$~\cite{ParticleDataGroup:2024cfk}. The recoil mass is
calculated using the reconstructed $\eta\to\gamma\gamma$ candidate as
$RM_\eta=\sqrt{(E_{\rm cm}-E_\eta)^2-|\vec{p}_\eta|^2}$,
where $E_{\rm cm}$ is the center-of-mass energy, while $E_{\eta}$ and $\vec{p}_{\eta}$ denote the energy and three-momentum of the $\eta$ candidate in the center-of-mass frame.
Using the same FOM optimization procedure,
the requirement $|RM_{\eta}-m_{J/\psi}|>0.007~\mathrm{GeV}/c^{2}$ is chosen.
In addition, contributions from
$\bar\Sigma^{0}\to\bar\Lambda\gamma$ are suppressed with a veto around the
$\bar\Sigma^{0}$ mass region. The photon not used in the selected
$\eta\to\gamma\gamma$ candidate is combined with the $\bar\Lambda$ candidate. 
The requirement
$|M_{\bar\Lambda\gamma}-m_{\Sigma^{0}}|>0.018~\mathrm{GeV}/c^{2}$
is optimized using the FOM procedure, where $m_{\Sigma^{0}}$ is the known $\Sigma^{0}$ mass~\cite{ParticleDataGroup:2024cfk}.

\section{Background analysis}
\label{sec:background}
Potential background contributions surviving the event selection are investigated using the inclusive $\psi(3686)$ MC sample. The decay chains of the candidate background events are identified with the
TopoAna package~\cite{Zhou:2020ksj}.
 The dominant decay chains can be categorized into three classes: (i) $\psi(3686)\to\eta J/\psi$ with $\eta\to\gamma\gamma$, (ii) $\psi(3686)\to\gamma\chi_{cJ}$ followed by $\chi_{cJ}\to\gamma J/\psi$ or hadronic decays, and (iii) multi-body direct decays possibly involving $K^*(892)$, $\Lambda$, or $\Sigma^*(1385)$ resonances. 

For the selected inclusive MC sample, the $M_{pK^{-}\bar{\Lambda}\eta}$ distribution exhibits a broad, non-peaking structure, indicating that the background 
predominantly originates from nonresonant processes and combinatorial contributions. 
A mild enhancement is found near $3.5~\mathrm{GeV}/c^{2}$, which is mainly attributed to 
cascade decays such as 
$\psi(3686)\!\to\!\gamma\chi_{c1}$ followed by 
$\chi_{c1}\!\to\!\gamma J/\psi$ with 
$J/\psi\!\to\!pK^{*-}(892)\bar{\Lambda}$ or 
$J/\psi\!\to\!pK^{-}\bar{\Sigma}^{*0}(1385)$. 
Dedicated exclusive MC studies show that the expected residual contributions from these two processes are 1 and 3 events, respectively. Therefore, their contributions are negligible and are not included in the final fit.

The possible quantum electrodynamics (QED) contribution from the process 
$e^{+}e^{-}\to\gamma\chi_{cJ}$ is evaluated with the continuum data at $\sqrt{s}=3.650$~GeV, corresponding to an integrated luminosity of $401.0$~pb$^{-1}$~\cite{BESIII:2024lks}. Applying the same event selection criteria as for the $\psi(3686)$ data, no surviving event is found. Hence, the QED background is negligible in this analysis.

\section{Branching fractions}
\label{sec:bf}

\begin{figure}[htbp]
  \centering
  \includegraphics[width=0.45\textwidth]{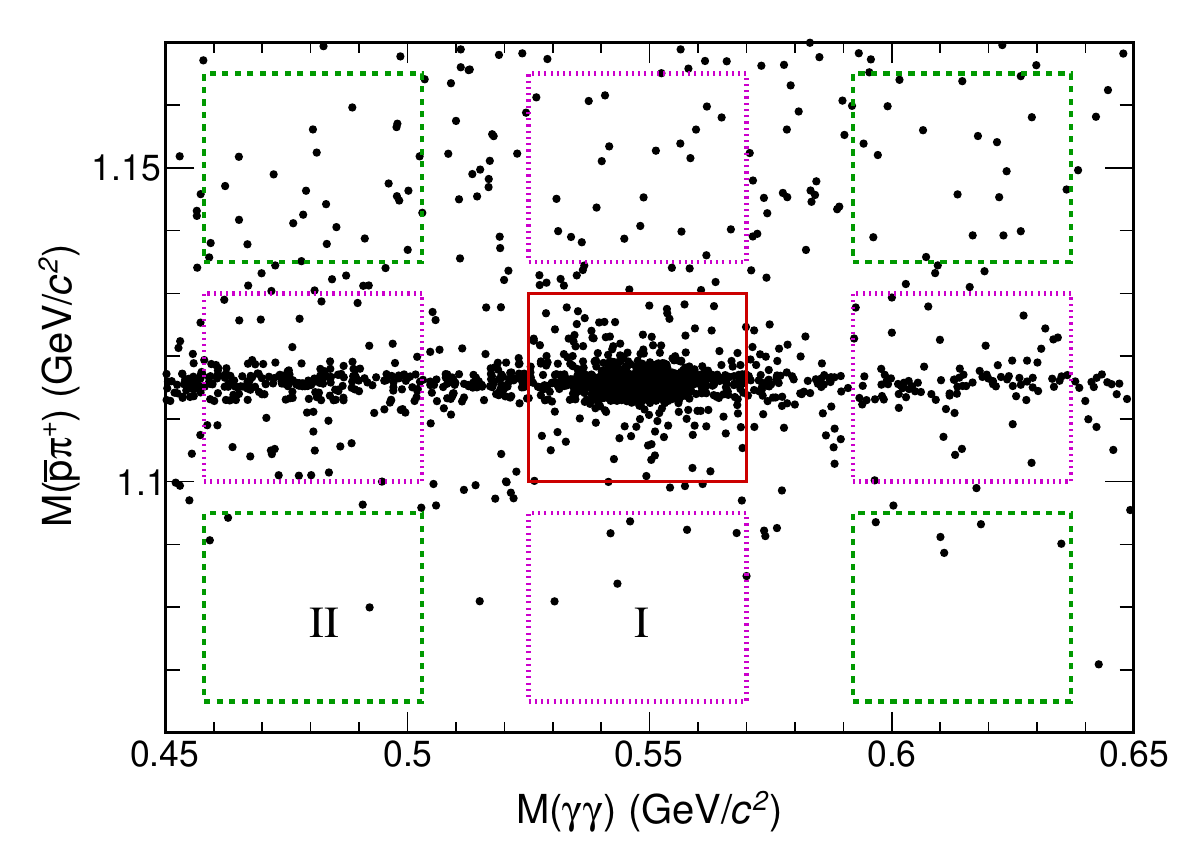}
  \caption{Two-dimensional distribution of $M_{\bar{p}\pi^{+}}$ versus $M_{\gamma\gamma}$ for the accepted events in data. 
The red solid rectangle denotes the $\eta$ signal region, 
the magenta dot-dashed rectangles denote the $\bar\Lambda$ and $\eta$ sideband~I regions, 
and the green dashed rectangles denote the $\bar\Lambda$ and $\eta$ sideband~II regions.
}

  \label{fig:sideband}
\end{figure}

To determine the signal yields, we perform an unbinned maximum-likelihood fit to the
$M_{pK^{-}\bar\Lambda\eta}$ spectrum of accepted candidates.
The fit is carried out simultaneously in the signal region and in the sideband regions, defined by events satisfying both the $\bar\Lambda$ and $\eta$ signal windows, and in the
$\bar\Lambda$ and $\eta$ sideband control samples shown in Fig.~\ref{fig:sideband}
(the sideband window definitions follow Sec.~\ref{sec:selection}).
This constrains backgrounds from misreconstructed $\bar\Lambda$ or $\eta$ candidates.

\begin{figure*}[htbp]
  \centering
  \includegraphics[width=5.5cm]{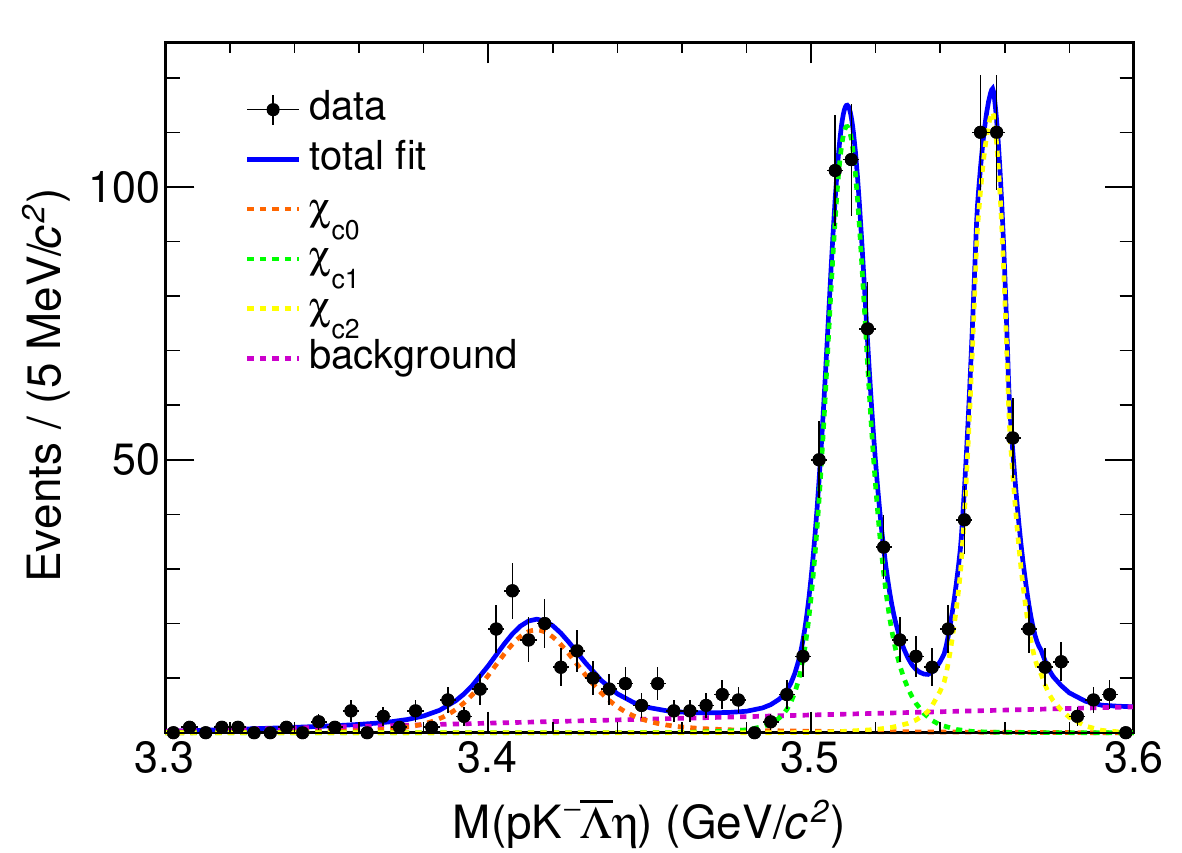}
  \includegraphics[width=5.5cm]{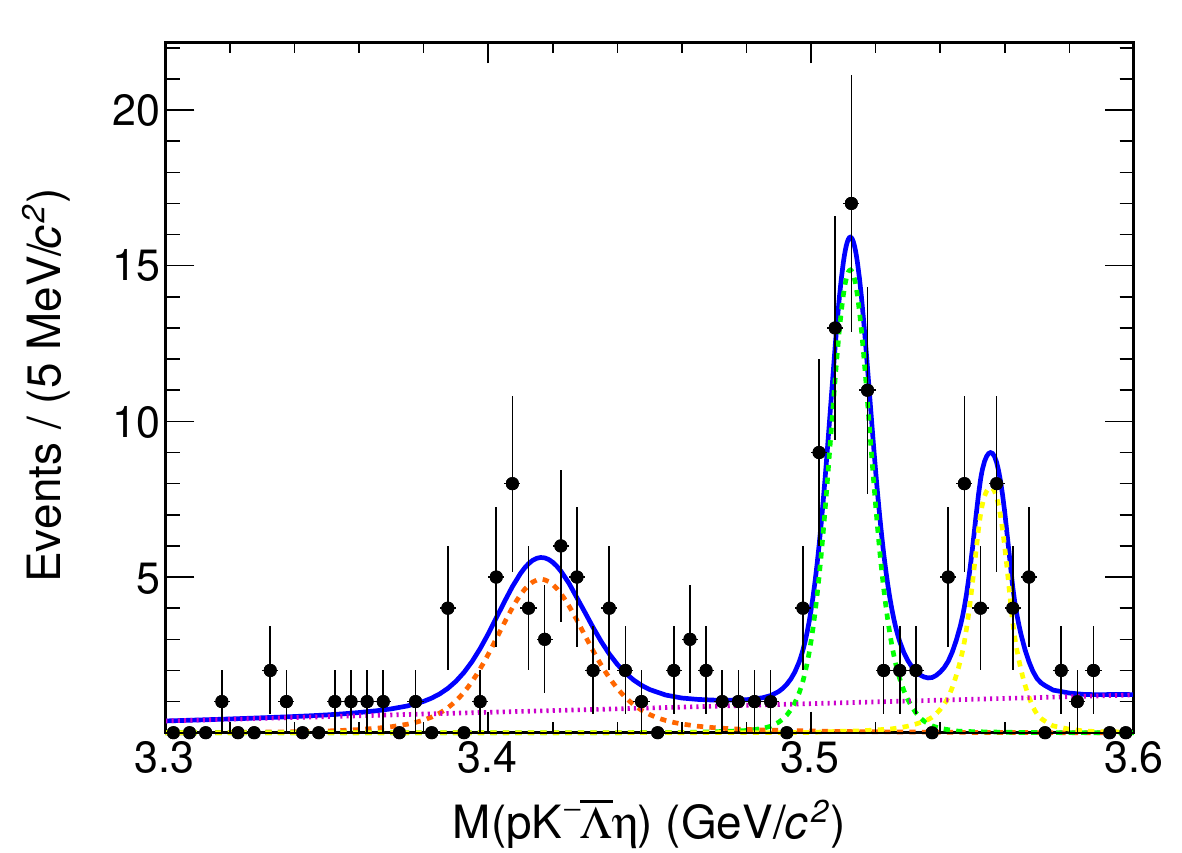}
  \includegraphics[width=5.5cm]{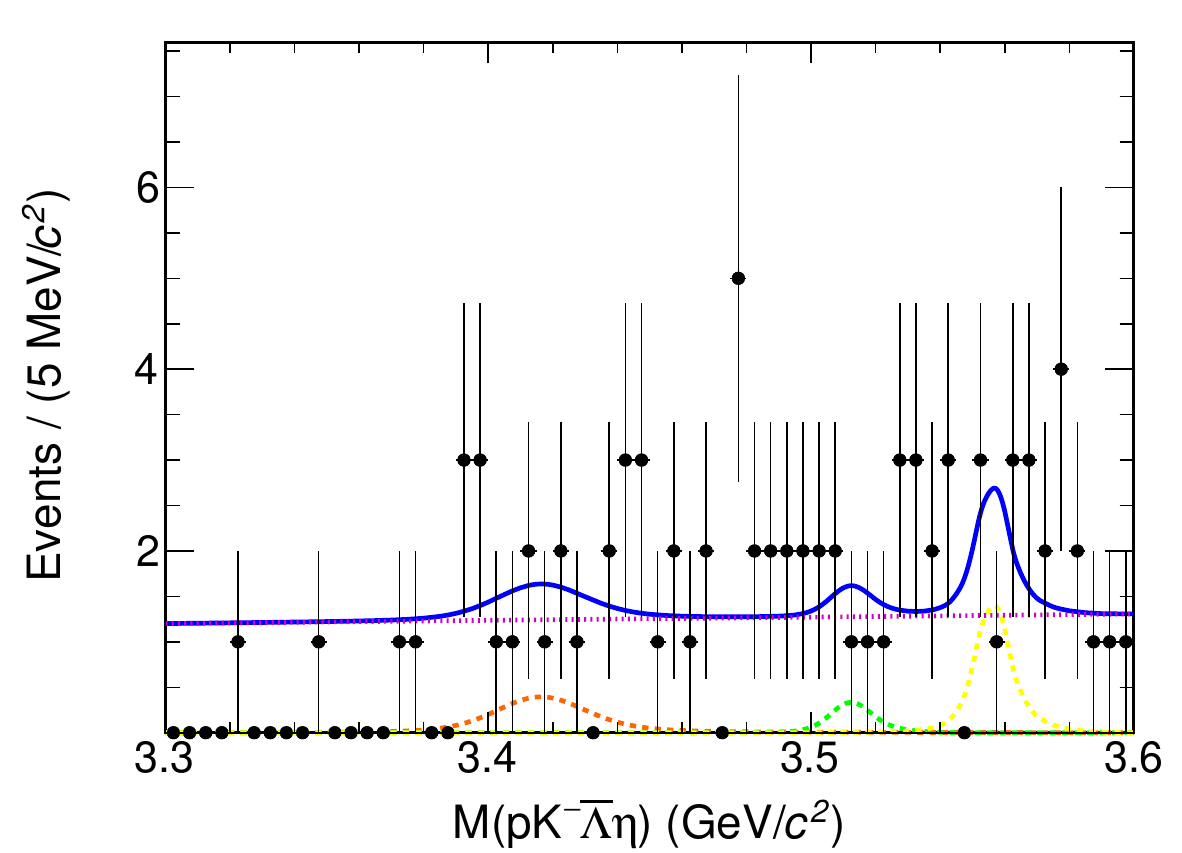}
  \caption{Fits to the $M_{pK^{-}\bar{\Lambda}\eta}$ distributions.
The left, middle, and right panels correspond to the signal,
sideband~I, and sideband~II regions, respectively.
Data are shown as black points with error bars. The blue solid curves represent the total fit results, the orange, green, and yellow dashed curves denote the individual $\chi_{c0}$, $\chi_{c1}$, and $\chi_{c2}$ signal components, respectively, and the magenta dotted curves denote the background components.
}

  \label{fig:fit}
\end{figure*}

In the simultaneous fit to $M_{pK^-\bar\Lambda\eta}$, the three $\chi_{cJ}$ peaks ($J=0,1,2$) are described by Breit--Wigner
functions with parameters determined from exclusive signal MC samples. Each Breit--Wigner function is convolved with a
single Gaussian function to account for the possible difference in mass
resolution between data and MC simulation. The Gaussian parameters are allowed
to float and are common between the signal and sideband samples for a given $J$.
The combinatorial background in both regions is described with a first-order
Chebyshev polynomial. Sideband normalizations are fixed by the geometric
scale factors determined from their window widths. The net signal yield obtained from the fitted $\chi_{cJ}$ peaking yields is
\( 
N_{\rm net}
= N_{\rm sig}
- \tfrac{1}{2} \cdot N_{\rm sideband~I}
+ \tfrac{1}{4} \cdot N_{\rm sideband~II}.
\)
Here, $N_{\rm sig}$ denotes the $\chi_{cJ}$ peaking yield in the signal region.
The net yields for each signal decay are  summarized in Table~\ref{tab:Branching}. The statistical significances are computed from the change in log-likelihood
with respect to the zero-signal hypothesis,
$\Delta(\ln\mathcal{L})=\ln\mathcal{L}_{\max}-\ln\mathcal{L}_{0}$. For each tested $\chi_{cJ}$ signal, the change in the number of degrees of freedom
is $\Delta n_{\rm dof}=3$, corresponding to the signal yield and the two Gaussian
smearing parameters, $\mu$ and $\sigma$. Statistical significances
exceeding $5\sigma$ are obtained for the $\chi_{c0}$, $\chi_{c1}$, and $\chi_{c2}$ signals.
Here, $n_{\rm dof}$ is the number of degrees of freedom. The projections of the
simultaneous fit in the signal as well as sideband~I and~II regions are shown
in Fig.~\ref{fig:fit}.

\begin{figure*}[htbp]
  \centering
  \includegraphics[width=\linewidth]{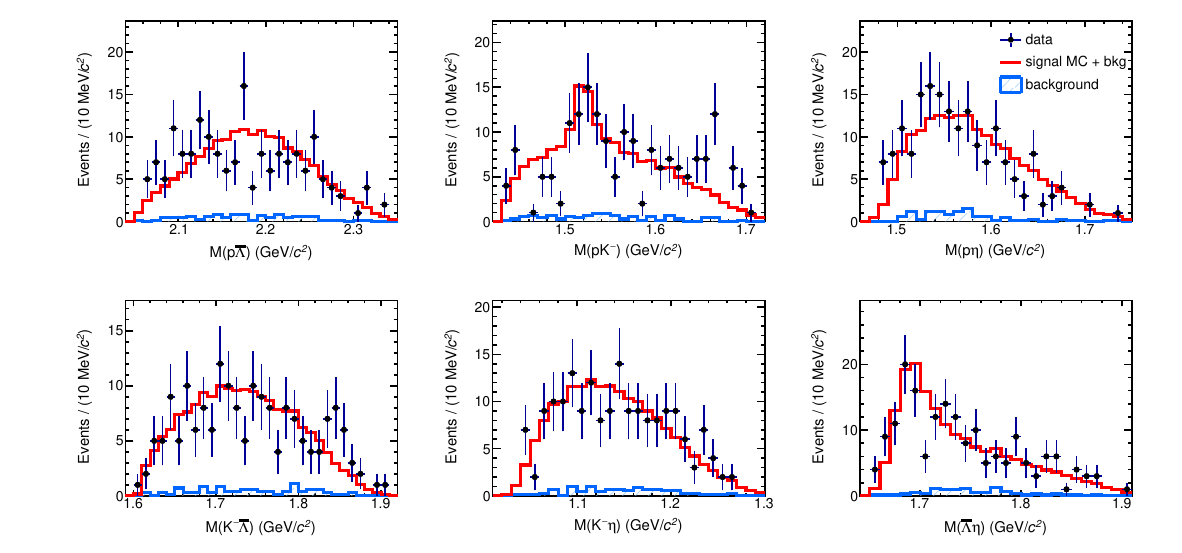} 
  \caption{Invariant mass distributions of the two-body combinations of the $\psi(3686)\to\gamma\chi_{c0}$, $\chi_{c0}\to pK^-\bar{\Lambda}\eta$ candidates.
The events are selected within the $\chi_{c0}$ signal region
of $[3.372,\,3.462]~\mathrm{GeV}/c^{2}$.
The data (points with error bars) are compared with the summed distribution of the signal MC including possible intermediate structures and the inclusive background (red solid histograms), and with the inclusive background (blue hatched histograms). All histograms are normalized to the data size.}
  \label{fig:masstwo_chic0}
\end{figure*}

\begin{figure*}[htbp]
  \centering
  \includegraphics[width=\linewidth]{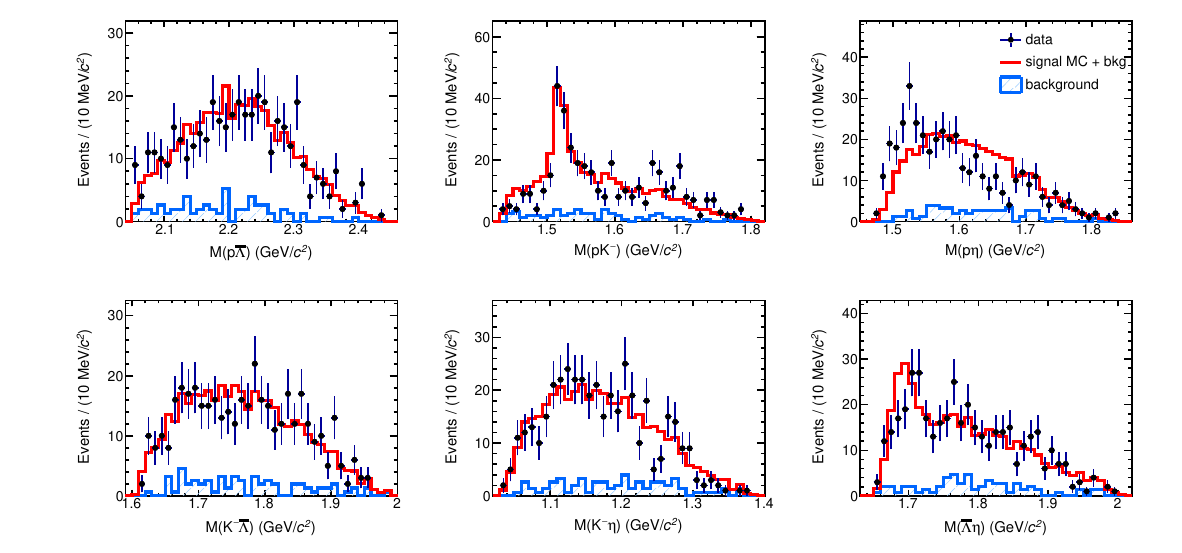} 
  \caption{Invariant mass distributions of the two-body combinations of the $\psi(3686)\to\gamma\chi_{c1}$, $\chi_{c1}\to pK^-\bar{\Lambda}\eta$ candidates.
The events are selected within the $\chi_{c1}$ signal region
of $[3.488,\,3.532]~\mathrm{GeV}/c^{2}$.
The data (points with error bars) are compared with the summed distribution of the signal MC including possible intermediate structures and the inclusive background (red solid histograms), and with the inclusive background (blue hatched histograms). All histograms are normalized to the data size.}
  \label{fig:masstwo_chic1}
\end{figure*}

\begin{figure*}[htbp]
  \centering
  \includegraphics[width=\linewidth]{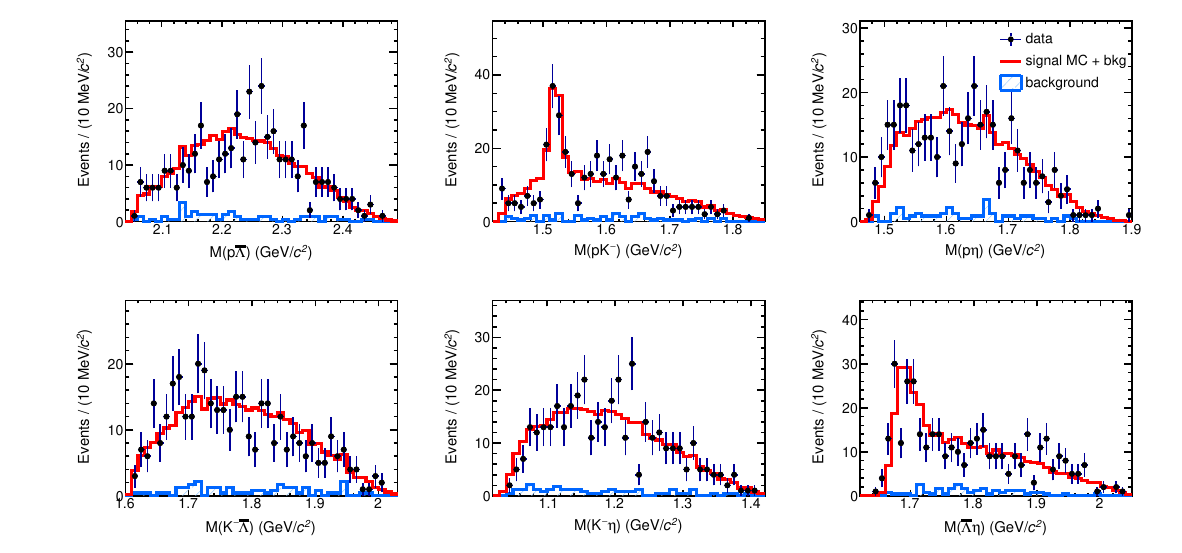} 
  \caption{Invariant mass distributions of the two-body combinations of the $\psi(3686)\to\gamma\chi_{c2}$, $\chi_{c2}\to pK^-\bar{\Lambda}\eta$ candidates.
The events are selected within the $\chi_{c2}$ signal region
of $[3.538,\,3.579]~\mathrm{GeV}/c^{2}$.
The data (points with error bars) are compared with the summed distribution of the signal MC including possible intermediate structures and the inclusive background (red solid histograms), and with the inclusive background (blue hatched histograms). All histograms are normalized to the data size.}
  \label{fig:masstwo_chic2}
\end{figure*}

As motivated by the invariant mass distributions of two-body comparison in
Figs.~\ref{fig:masstwo_chic0}--\ref{fig:masstwo_chic2},
enhancements in $M_{pK^-}$ and $M_{\bar\Lambda\eta}$ indicate possible contributions
from intermediate hyperon resonances $\Lambda(1520)$ and $\bar\Lambda(1690)$.
To account for their impact on the detection efficiency, a mixed signal MC sample is constructed
by combining the pure PHSP process with these resonant sub-processes,
weighted according to their relative fractions extracted from data.
The resulting detection efficiencies are
3.7\%, 5.5\%, and 5.7\% for $\chi_{c0}$, $\chi_{c1}$, and $\chi_{c2}$, respectively.

Finally, the product branching fractions for the cascade decays
$\psi(3686)\!\to\!\gamma\chi_{cJ}$ and
$\chi_{cJ}\!\to\!pK^{-}\bar{\Lambda}\eta$ are determined by
\begin{equation}
\label{eq:prodBF}
\begin{aligned}
\mathcal{B}\!\big(\psi(3686)\!\to\!\gamma\chi_{cJ}\big)\,
\mathcal{B}\!\big(\chi_{cJ}\!\to\!pK^{-}\bar{\Lambda}\eta\big)
\\[2pt]
= \frac{N_{\rm net}}{\,N_{\psi(3686)}\,\varepsilon_{J}\,
\mathcal{B}\!\big(\bar{\Lambda}\!\to\!\bar p\pi^{+}\big)\,
\mathcal{B}\!\big(\eta\!\to\!\gamma\gamma\big)}\,,
\end{aligned}
\end{equation}
where $N_{\psi(3686)}=(2712.4\pm14.3)\times10^{6}$, and
$\varepsilon_{J}$ is the mixed-signal MC efficiency for a given $J$.
The secondary branching fractions are taken from the PDG:
$\mathcal{B}(\psi(3686)\!\to\!\gamma\chi_{c0,1,2})=(9.75\pm0.22)\%$, $(9.75\pm0.27)\%$, and $(9.38\pm0.23)\%$,
$\mathcal{B}(\bar{\Lambda}\!\to\!\bar p\pi^{+})=(64.1\pm0.5)\%$,
and $\mathcal{B}(\eta\!\to\!\gamma\gamma)=(39.36\pm0.18)\%$.
The resulting product branching fractions and the corresponding
$\mathcal{B}(\chi_{cJ}\!\to\!pK^{-}\bar{\Lambda}\eta)$, including charge-conjugate modes,
are listed in Table~\ref{tab:Branching}.

\begin{table*}[htbp]
  \centering
  \renewcommand{\arraystretch}{1.2}
  \caption{Signal yields, efficiencies, and branching-fraction results for
  $\chi_{cJ}\!\to\!pK^{-}\bar\Lambda\eta$ (statistical uncertainties only).}
  \label{tab:Branching}
  \begin{tabular}{cccc}
    \hline\hline
    & $\chi_{c0}$ & $\chi_{c1}$ & $\chi_{c2}$ \\
    \hline
    $N_{\rm net}$ & $130.2\pm16.8$ & $359.7\pm21.8$ & $341.2\pm21.2$ \\
    $\varepsilon_J$ (\%) & $3.7$ & $5.5$ & $5.7$ \\
    $\mathcal{B}\big(\psi(3686)\!\to\!\gamma\chi_{cJ}\big)\,
     \mathcal{B}\big(\chi_{cJ}\!\to\!pK^{-}\bar\Lambda\eta\big)$ ($\!\times\!10^{-7}$)
     & $51.4\pm6.6$ & $95.6\pm5.8$ & $87.5\pm5.4$ \\
    $\mathcal{B}\big(\chi_{cJ}\!\to\!pK^{-}\bar\Lambda\eta\big)$ ($\!\times\!10^{-5}$)
     & $5.3\pm0.7$ & $9.8\pm0.6$ & $9.3\pm0.6$ \\
    \hline\hline
  \end{tabular}
\end{table*}

\section{SYSTEMATIC UNCERTAINTY}
\label{sec:systematics}

The systematic uncertainties on the branching fraction measurements arise from multiple sources. 
These include the uncertainties associated with the total number of $\psi(3686)$ events, tracking and PID efficiencies, photon detection, $\eta$ mass-window requirement, $\bar\Lambda$ reconstruction, the $\pi^{0}$, $J/\psi$, and $\Sigma^{0}$ vetoes, the modelling of combinatorial background shapes, the choices of fit ranges and sideband regions, the 4C kinematic fit, and the limited signal-MC statistics. 
In addition, the uncertainties of the external branching fractions quoted from the PDG are included. 
Possible contributions from the $\Lambda(1670)$ and $N(1535)$ structures are also evaluated and incorporated as additional systematic uncertainties.  
All these sources are treated as independent and summarized in Table~\ref{tab:syst}, with their individual evaluations described in the following paragraphs. Here, a dash in Table~\ref{tab:syst} denotes that the corresponding source is not applicable 
to the given $\chi_{cJ}$ mode.

The total number of $\psi(3686)$ events is
$N_{\psi(3686)}=(2712.4\pm14.3)\times 10^{6}$, determined with inclusive
hadronic events.  The associated uncertainty is $0.5\%$~\cite{BESIII:2024lks}.

 The tracking and PID efficiencies are studied with the control samples of $J/\psi\to K^*(892)K$~\cite{Yuan:2015wga} and $J/\psi\to p\bar p\pi^{+}\pi^{-}$~\cite{BESIII:2023pqp}.
The systematic uncertainty due to tracking or PID is assigned as $1.0\%$ per charged track.
Since the tracking effects of  $\bar{p}$ and ${\pi}^{+}$ have been considered in the estimation of the systematic uncertainty in $\bar\Lambda$ reconstruction, only the proton and kaon are considered here. The total  uncertainties are $2.0\%$ for tracking of two charged tracks and $2.0\%$ for PID of two charged tracks. 

The systematic uncertainty due to the photon detection is estimated with the control sample of $J/\psi\!\to\!\pi^{+}\pi^{-}\pi^{0}$~\cite{BESIII:2015rug} to be $1.0\%$ per photon. Three photons in the final state lead the total uncertainty to be $3.0\%$.

The uncertainty due to the $\eta$ mass–window requirement is evaluated
with $\psi(3686)\!\to\!\eta J/\psi$, $J/\psi\!\to\!\ell^{+}\ell^{-}$
($\ell=e,\mu$).  The change of signal efficiency after smearing the $\eta$ signal shape by a Gaussian function with parameters derived  from the control sample, $1.0\%$~\cite{BESIII:2022rsg} is taken as the systematic uncertainty.

The systematic uncertainties in the $\bar\Lambda$ reconstruction are studied with the control samples of $J/\psi\!\to\!pK^{-}\bar\Lambda$
and $J/\psi\!\to\!\Lambda\bar\Lambda$~\cite{BESIII:2018ciw}. The differences in $\bar\Lambda$  reconstruction efficiencies, including the $\bar{p}\pi^{+}$ tracking,
decay-length, mass-window, vertex-fit, and secondary-vertex selections are taken as the systematic uncertainties due to $\bar\Lambda$ reconstruction, which are $2.7\%$ for $\chi_{c0}$, $1.2\%$ for $\chi_{c1}$, and $0.8\%$ for $\chi_{c2}$~\cite{BESIII:2022rsg}.

The systematic uncertainties of possible mismodelling of the veto requirements are examined by varying the $\pi^{0}$, $J/\psi$, and $\Sigma^{0}$ veto windows around their nominal definitions by $\pm1~\mathrm{MeV}$. The varied requirements are applied to both data and signal MC samples, and the changes in the efficiency-corrected signal yields are taken as the systematic uncertainties, which are
$(2.2,\,0.9,\,0.6)\%$ for the $\pi^{0}$ veto,
$(0.1,\,0.5,\,3.1)\%$ for the $J/\psi$ veto, and
$(0.9,\,0.2,\,0.1)\%$ for the $\Sigma^{0}$ veto, for $(\chi_{c0},\,\chi_{c1},\,\chi_{c2})$, respectively.

The systematic uncertainties of the combinatorial background shapes are estimated by varying the nominal first-order Chebyshev polynomial with several alternative background parameterizations in the signal, Sideband I, and Sideband II regions simultaneously. The alternative models include a second-order Chebyshev polynomial, a threshold-like function $(x-x_{0})^{\alpha}$, $(1+a x)(x-x_{0})^{\alpha}$, and $e^{a x}(x-x_{0})^{\alpha}$, where $x_{0}$ is fixed to the threshold value and the other parameters are allowed to float. The largest deviations are obtained with the $(x-x_{0})^{\alpha}$ parameterization, and are taken as the systematic uncertainties due to the background shape, corresponding to $5.1\%$, $0.8\%$, and $0.5\%$ for $\chi_{c0}$, $\chi_{c1}$, and $\chi_{c2}$, respectively.

The systematic uncertainties of the fit range are examined by several alternative fit intervals, $[3.30,3.61]$, $[3.30,3.59]$, $[3.29,3.60]$,
and $[3.31,3.60]~\text{GeV}/c^{2}$. The yield variations are consistent with
statistical fluctuations ($<2\sigma$), therefore, no additional systematic uncertainty is assigned.

The systematic uncertainties of the $\bar\Lambda$ and $\eta$ sideband windows are estimated by varying each window by $\pm 1\sigma$. The changes of the net signal yields, $1.0\%$, $0.6\%$, and $0.7\%$ for
$\chi_{c0}$, $\chi_{c1}$, and $\chi_{c2}$, are taken as the systematic uncertainties. 

The systematic uncertainties of the 4C kinematic fit are assigned as the differences between the detection efficiencies before and after the helix parameter corrections~\cite{BESIII:2012mpj}, which are $1.3\%$, $1.2\%$, and $1.3\%$ for $\chi_{c0}$, $\chi_{c1}$, and $\chi_{c2}$.

The uncertainties from limited MC statistics are calculated from
$\sqrt{\frac{1-\epsilon}{N\cdot\epsilon}}$,
where $\epsilon$ is the detection efficiency and $N$ is the number of signal MC events. They are $0.2\%$ for all channels.

The uncertainties of the branching fractions quoted from the PDG are also considered: $(2.3,~2.8,~2.5)\%$ for $\mathcal{B}(\psi(3686)\!\to\!\gamma\chi_{c0,1,2})$,~0.8\% for $\mathcal{B}(\bar\Lambda\!\to\!\bar{p}\pi^{+})$ and 0.5\% for 
$\mathcal{B}(\eta\!\to\!\gamma\gamma)$.

As seen in Fig.~\ref{fig:masstwo_chic0}, a mild enhancement is observed near $1.67~\mathrm{GeV}/c^{2}$ in the $M_{pK^-}$ distribution of the $\chi_{c0}$ sample, which may be due to the $\Lambda(1670)$ resonance. As seen in Fig.~\ref{fig:masstwo_chic1}, a mild enhancement is observed near $1.535~\mathrm{GeV}/c^{2}$ in the $M_{p\eta}$ distribution of the $\chi_{c1}$ sample, which may be due to the $N(1535)$ resonance. Due to the insufficient significances, their effects on the branching fraction measurements are considered as systematic uncertainties. Reweighting the $\chi_{c0}$ and $\chi_{c1}$ signal MC events to the corresponding $M_{pK^-}$ and $M_{p\eta}$ line shapes and repeating the full selections yield efficiency shifts of $2.7\%$ and $2.8\%$, which are assigned as additional uncertainties for the $\chi_{c0}$ and $\chi_{c1}$ decays, respectively.

For each signal decay, the total systematic uncertainty is calculated by adding systematic uncertainties quadratically under the assumption that all sources are independent.

\begin{table}[t]
\centering
\caption{Relative systematic uncertainties (in \%) on 
$\mathcal{B}(\chi_{cJ}\to pK^{-}\bar{\Lambda}\eta)$. 
A dash indicates that the corresponding systematic uncertainty is not applicable for that $\chi_{cJ}$ mode.}
\label{tab:syst}
\begin{tabular}{lccc}
\hline\hline
Source & $\chi_{c0}$ & $\chi_{c1}$ & $\chi_{c2}$ \\
\hline
$N_{\psi(3686)}$                    & 0.5 & 0.5 & 0.5\\
		Tracking                            &2.0 &2.0 &2.0\\
		PID                                 &2.0 &2.0 &2.0\\
		$\gamma$ selection                  & 3.0 & 3.0 & 3.0\\
		$\eta$ mass window                 & 1.0 & 1.0 & 1.0\\
		$\bar \Lambda$ reconstruction      & 2.7 &1.2 &0.8\\
        $\pi^0$ veto  & 2.2 & 0.9 & 0.6 \\
        $J/\psi$ veto & 0.1 & 0.5 & 3.1 \\
        $\Sigma^0$ veto & 0.9 & 0.2 & 0.1 \\
		Background shape                 & 5.1 & 0.8 & 0.5\\
        Sideband regions  
           & 1.0 & 0.6 & 0.7\\
		4C kinematic fit                    & 1.3 & 1.2 & 1.3 \\
		MC statistics            & 0.2 & 0.2 &0.2 \\
        $\mathcal{B}(\psi(3686) \to \gamma\chi_{cJ})$ & 2.3& 2.8& 2.5\\
		$\mathcal{B}(\bar{\Lambda} \to \bar{p}\pi^+)$& 0.8& 0.8 & 0.8\\
		$\mathcal{B}(\eta\to\gamma\gamma)$    & 0.5 & 0.5 & 0.5\\
        $\Lambda(1670)$ for $\chi_{c0}$
        & 2.7 & - & -\\
        $N(1535)$ for $\chi_{c1}$
        & - & 2.8 & -\\
        \hline
		Total                   &8.6  & 6.3 & 6.2\\
\hline\hline
\end{tabular}
\end{table}

\section{Summary}
Using a sample of $(2712.4 \pm 14.3)\times 10^{6}$ $\psi(3686)$ events collected
with the BESIII detector, the decays $\chi_{cJ} \to pK^- \bar{\Lambda}\eta$
($J=0,1,2$) are observed for the first time. The statistical significances
of the $\chi_{c0}$, $\chi_{c1}$, and $\chi_{c2}$ signals all exceed $5\sigma$.
The measured branching fractions are
\[
\begin{aligned}
\mathcal{B}(\chi_{c0}\!\to\! pK^{-}\bar\Lambda\eta)
  &= (5.3\pm0.7\pm0.5)\times10^{-5},\\
\mathcal{B}(\chi_{c1}\!\to\! pK^{-}\bar\Lambda\eta)
  &= (9.8\pm0.6\pm0.6)\times10^{-5},\\
\mathcal{B}(\chi_{c2}\!\to\! pK^{-}\bar\Lambda\eta)
  &= (9.3\pm0.6\pm0.6)\times10^{-5},
\end{aligned}
\]
where the first and second uncertainties are statistical and systematic, respectively.
To improve the determination of the detection efficiencies and account for possible hadronic
substructures, a mixed signal MC sample is constructed by combining the PHSP process
with components used to describe the observed structures in the $pK^-$ and
$\bar{\Lambda}\eta$ invariant mass distributions. These structures are consistent with
$\Lambda(1520)$ and $\bar{\Lambda}(1690)$, respectively. The mixed signal MC samples
provide a more representative description of the selected signal events and are used to
determine the detection efficiencies for the branching-fraction measurements of
$\chi_{cJ}\to pK^-\bar{\Lambda}\eta+c.c.$. The results obtained in this work enrich the experimental knowledge of
$P$-wave charmonium decays involving baryon--antibaryon--meson systems~\cite{BESIII:2022rsg}, and provide
useful input for understanding baryon production and hadronization mechanisms in the
nonperturbative QCD regime.

\textbf{Acknowledgement}

The BESIII Collaboration thanks the staff of BEPCII (https://cstr.cn/31109.02.BEPC) and the IHEP computing center for their strong support. This work is supported in part by National Key R\&D Program of China under Contracts Nos. 
2025YFA1613900,
2023YFA1606000, 2023YFA1606704; National Natural Science Foundation of China (NSFC) under Contracts Nos. 11635010, 11935015, 11935016, 11935018, 12025502, 12035009, 12035013, 12061131003, 12192260, 12192261, 12192262, 12192263, 12192264, 12192265, 12221005, 12225509, 12235017, 12342502, 12361141819, 12535005; the Chinese Academy of Sciences (CAS) Large-Scale Scientific Facility Program; the Strategic Priority Research Program of Chinese Academy of Sciences under Contract No. XDA0480600; CAS under Contract No. YSBR-101; 100 Talents Program of CAS; The Institute of Nuclear and Particle Physics (INPAC) and Shanghai Key Laboratory for Particle Physics and Cosmology; Agencia Nacional de Investigación y Desarrollo de Chile (ANID), Chile under Contract No. ANID CCTVal CIA250027; ERC under Contract No. 758462; German Research Foundation DFG under Contract No. FOR5327; Istituto Nazionale di Fisica Nucleare, Italy; Knut and Alice Wallenberg Foundation under Contracts Nos. 2021.0174, 2021.0299, 2023.0315; Ministry of Development of Turkey under Contract No. DPT2006K-120470; National Research Foundation of Korea under Contract No. RS-2026-25486791; National Science and Technology fund of Mongolia; Polish National Science Centre under Contract No. 2024/53/B/ST2/00975; STFC (United Kingdom); Swedish Research Council under Contract No. 2019.04595; U. S. Department of Energy under Contract No. DE-FG02-05ER41374


\bibliography{ref}

\end{document}